\documentstyle[epsfig]{sup2}

\vspace{-8pt}

\title[Superlattices and Microstructures, Vol.\ ??, No.\ ??, 1999]
{ Contact resistance of quantum tubes }

\author[Superlattices and Microstructures, Vol.\ ??, No.\ ??, 1999]
{Niels Asger Mortensen$^*$, Kristinn Johnsen$^{\dag *}$,\cr\vspace{8pt} Antti-Pekka
  Jauho$^*$, Karsten Flensberg$^\ddag$ \cr
{\normalsize\it $^*$Mikroelektronik Centret, Technical University of
  Denmark,}\cr {\normalsize\it \O rsteds Plads, Bld. 345 east, DK-2800 Kgs. Lyngby,
  Denmark}\cr{\normalsize\it $^\dag$Nordic Institute for
    Theoretical Physics, Blegdamsvej 17, DK-2100 Copenhagen, Denmark
    \O}\cr{\normalsize\it $^\ddag$\O rsted Laboratory, Niels Bohr Institute, University of Copenhagen,}\cr{\normalsize\it Universitetsparken 5, DK-2100 Copenhagen \O, Denmark}\cr
}

\begin{document}
\label{firstpage}
\maketitle
\sloppy
\begin{center}
\received{(Submitted \today)}
\end{center}

\begin{abstract}
  We consider the conductance of a quantum tube connected to a
  metallic contact. The number of angular momentum states that the
  tube can support depends on the strength of the radial confinement.
  We calculate the transmission coefficients which yield the
  conductance via the Landauer formula, and discuss the relation of
  our results to armchair carbon nanotubes embedded in a metal. For Al
  and Au contacts and tubes with a realistic radial confinement we
  find that the transmission can be close to unity corresponding to a
  contact resistance close to $h/2e^{2}$ per band at the Fermi level
  in the carbon nanotube.
\end{abstract}

\section{Introduction}
 
Since the recent discovery of the carbon nanotubes by Iijima
\cite{IIJIMA91} there has been a significant progress \cite{DEKKER99}
in the studies of the conducting properties of both single-walled
\cite{TANS97} and multi-walled \cite{EBBESEN96} carbon nanotubes.
Conductance of a mesoscopic system connected to metallic reservoirs is
well understood and is usually described by the Landauer formula
\cite{LANDAUER}.  For quantum point contacts in semiconductor
structures and in metallic nanowires it is well establish
experimentally that the differential conductance to a good
approximation is quantized in units of $G_{0}=2e^{2}/h$ and at zero
temperature given by $G=G_{0}N$ where $N$ is the number of propagating
modes. In carbon nanotubes with metallic contacts most experiments
show that the conductance is less than the conductance which one
should expect for a smooth interface between tube and metal, e.g.
$G=4e^{2}/h$ for metallic single-walled tubes where the extra factor
of 2 comes from the two $\pi$ bands that are crossing the Fermi level
\cite{twobands}. The reasons for this lower conductance are still not
fully known. Theoretically, several groups have considered the effects
of vacancies \cite{CHICO96}, disorder \cite{disorder}, distortion
\cite{ROCHEFORT}, and doping \cite{FARAJIAN} on the conductance of
carbon nanotubes. The conducting properties have also been studied in
e.g. the context of junctions between different metallic carbon
nanotubes \cite{TAMURA}, Aharonow--Bohm effect in the presence of a
magnetic field \cite {ANDO}, and the Luttinger liquid behavior of a
one-dimensional gas of interacting electrons \cite{LUTTINGER}. Also
the ideal ``hollow quantum cylinder'', i.e. a two-dimensional electron
gas on a cylinder, has been studied in context of the difference
between strip-like wires and tubes \cite{CHAPLIK98}. However, with the
exception of the recent qualitative study of Tersoff \cite{TERSOFF99}
and the recent modeling-works of Anantram {\it et al.}
\cite{ANANTRAM99} and Sanvito {\it et al.} \cite{SANVITO99},
less attention has been focused on the conditions for a good
transmission between tube and a metal contact which is an important
issue for practical devices with carbon nanotubes, or other quantum tubes.

In quantum point contacts an adiabatic interface between the wire and
reservoirs ensures a transmission coefficient close to unity
\cite{glazman}. The condition for adiabaticity is that the shape of
the contact region varies slowly on the scale of the Fermi wave
length. In the opposite case with an abrupt interface, i.e.
quasi-one-dimensional lead connected to a wide two-dimensional
contact, Szafer and Stone \cite{SZAFER89} found that the transmission
rapidly increases to unity as the width of the confined region exceeds
half of the Fermi wavelength, thus giving a reflectionless contact.

For the contact between a quantum tube and a three-dimensional metal
it is not obvious that the assumption of an ideal reflectionless
contact applies and the aim of this work is to study the contact
resistance for this case.

\begin{figure}[htb]
\begin{center}
\epsfig{file=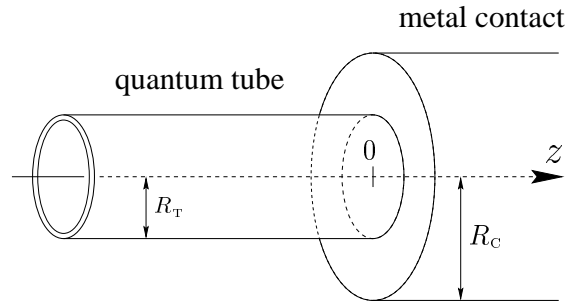, width=0.49\columnwidth}
\end{center}
\caption{Contact between a quantum tube ($z<0$) of radius
  $R_{\scriptscriptstyle\rm T}$ and a three-dimensional cylindrical
  free-electron metallic wire ($z>0$) of radius
  $R_{\scriptscriptstyle\rm C}\gg R_{\scriptscriptstyle\rm T}$.}
\label{FIG1}
\end{figure}

The model we are studying is that of a hollow quantum cylinder of
radius $R_{\scriptscriptstyle\rm T}$ contacted by a three-dimensional
free-electron metal which we for convenience model by a cylindrical
wire with radius $R_{\scriptscriptstyle\rm C}\gg
R_{\scriptscriptstyle\rm T}$, see Fig. \ref{FIG1}. The system thus has
full cylindrical symmetry and the angular momentum quantum number $m$
can therefore be used to label the scattering states.

For the coupling of the quantum tube to the contact it is necessary to
take a radial confinement potential for the quantum tube into account
and here we model the confinement by an attractive delta-function
potential. As an example we apply this model to metal contacts of Al
or Au; the quantum tube parameters are chosen to mimic armchair carbon
nanotubes: the
strength of the confinement can be related to the work function for
the material that constitutes the tube and in the case of a carbon
nanotube we relate it to the work function of graphene. It should be
noted that the employed free electron model does not fully describe
the actual band structure of carbon nanotubes. Nevertheless, a study
of contact resistance within this idealized model should yield
valuable insights which are relevant to real materials.

The paper is organized as follows: In Section II the eigenstates of a
quantum tube connected to a cylindrical metal contact are found. In
Section III these eigenstates are used to construct the scattering
states to find the transmission coefficient, and hence the conductance
of the contact. In Section IV, we apply our model to contacts between
an armchair carbon nanotube and a metal. Finally, in Section V
discussion and conclusions are given. Essential details of analytical
calculations are given in Appendices A and B.

\section{The eigenstates}

We separate the discussion into two parts: first we find the eigenstates
in the tubular geometry and then the eigenstates for the cylindrical
metal contact. In Section III the matching of these eigenstates are
used to construct the scattering states of the contact.

\subsection{Quantum tube}

The quantum tube of radius $R_{\scriptscriptstyle\rm T}$ with
otherwise free electrons is modeled by the Hamiltonian

\begin{equation}
\hat{{\cal {H}}}_{\scriptscriptstyle\rm T}
=-\frac{\hbar ^{2}}{2m_e}\left[ \frac{\partial ^{2}}{\partial z^{2}}
+\frac{\partial ^{2}}{\partial r^{2}}
+\frac{1}{r}\frac{\partial }{\partial r}
+\frac{1}{r^{2}}\frac{\partial ^{2}}{\partial \phi ^{2}}\right]
+V_{\scriptscriptstyle\rm T}(r), 
\end{equation}
with a confining potential given by an attractive delta function potential 
\begin{equation}
V_{\scriptscriptstyle\rm T}(r)=-H\delta (r-R_{\scriptscriptstyle\rm T}),
\end{equation}
where the confinement strength $H$ is taken positive.

The eigenstates of the Schr\"{o}dinger equation have the form 
\begin{equation}
\Psi _{m}(r,\phi,z)=R_{m}(r)\chi _{m}(\phi )\psi _{m}(z),
\end{equation}
with angular and longitudinal wave functions 
\begin{eqnarray}
\chi _{m}(\phi ) &=&(2\pi )^{-1/2}\exp [im\phi ], \\
\psi _{m}(z) &=&\left[ k_{m}(E)\right]^{-1/2}
\exp \left[ \pm ik_{m}(E)z \right] ,
\end{eqnarray}
where the angular momentum quantum numbers $m$ are integers,
$k_{m}(E)=\left[ \frac{2m_e}{\hbar ^{2}}\left( E-\varepsilon
    _{m}\right) \right] ^{1/2}$ is the wave vector associated to the
longitudinal free propagation, and $E=E_m+\varepsilon_m$ is the total
energy of the state. Here, $E_m>0$ is the energy associated to the
longitudinal propagation and $\varepsilon _{m}<0$ is the (binding)
energy associated to the transverse motion. We can relate the strength
of the confinement to the work function $W=\left|\varepsilon
  _{m}\right|-\hbar^2k_{{\rm F}}^2/2m_e$, which is the energy required
to remove an electron at the Fermi level (disregarding surface charge
effects). The normalization $\left[ k_{m}(E)\right] ^{-1/2}$ is chosen
such that the propagating modes carry the same amount of current.

The radial wave function $R_{m}(r)$ satisfies 
\begin{equation}\label{Schrodinger_radial}
\left\{ r^{2}\frac{\partial ^{2}}{\partial r^{2}}
+r\frac{\partial }{\partial r}
+\left[ \frac{2m_e\varepsilon _{m}}{\hbar ^{2}}r^{2}-m^{2}\right]
\right\} R_{m}(r)=
-\gamma R_{\scriptscriptstyle\rm T} 
\delta(r-R_{\scriptscriptstyle\rm T})R_{m}(r),
\end{equation}
where $\gamma \equiv 2m_e HR_{\scriptscriptstyle\rm T}/\hbar ^{2}$ is
a dimensionless confinement strength. For the bound states
($\varepsilon_m <0$) and $r\neq R_{\scriptscriptstyle\rm T}$ this
equation has the form of Bessel's modified differential equation \cite
{OLIVER}. The solutions are given by modified Bessel functions of
order $m$ of the first and second kind, so that the full solution is
given by
\begin{eqnarray}
R_{m}(r) &=& \left\{ \begin{array}{ccc}
A_{m}I_{m}(\kappa _{m}r)&,&r<R_{\scriptscriptstyle\rm T}\\ 
B_{m}K_{m}(\kappa _{m}r)&,&r> R_{\scriptscriptstyle\rm T} 
\end{array}\right.,
\end{eqnarray}
where $\kappa _{m}\equiv \left[ 2m_e\left| \varepsilon _{m}\right|
  /\hbar ^{2}\right] ^{1/2}$. At $r=R_{\scriptscriptstyle\rm T}$, the
radial wave function is continuous and the appropriate matching
condition for the derivative $\partial R_{m}(r)/\partial r$ at
$r=R_{\scriptscriptstyle\rm T}$ is found by integrating Eq.
(\ref{Schrodinger_radial}) from $R_{\scriptscriptstyle\rm
  T}^{-}=R_{\scriptscriptstyle\rm T}-0^+ $ to
$R_{\scriptscriptstyle\rm T}^{+}=R_{\scriptscriptstyle\rm T}+0^+ $. In
this way the matching conditions become

\begin{eqnarray}
R_{m}(R_{\scriptscriptstyle\rm T}^{+})
-R_{m}(R_{\scriptscriptstyle\rm T}^{-}) 
&=&0, \\
\left. \frac{\partial R_{m}(r)}{\partial r}\right|
_{R_{\scriptscriptstyle\rm T}^{+}}
-\left. \frac{\partial R_{m}(r)}{\partial r}\right|
_{R_{\scriptscriptstyle\rm T}^{-}} 
&=&-\gamma \frac{R_{m}(R_{\scriptscriptstyle\rm T})}
{R_{\scriptscriptstyle\rm T}},
\end{eqnarray}
and we get the following equation for the normalization coefficients
\begin{equation}
\left( 
\begin{array}{cc}
I_{m}(\kappa _{m}R_{\scriptscriptstyle\rm T}) 
& -K_{m}(\kappa _{m}R_{\scriptscriptstyle\rm T}) \\ 
I_{m-1}(\kappa _{m}R_{\scriptscriptstyle\rm T})
+I_{m+1}(\kappa _{m}R_{\scriptscriptstyle\rm T})
-\frac{2\gamma I_{m}(\kappa_{m}R_{\scriptscriptstyle\rm T})}
{\kappa _{m}R_{\scriptscriptstyle\rm T}} 
& K_{m-1}(\kappa _{m}R_{\scriptscriptstyle\rm T})
+K_{m+1}(\kappa _{m}R_{\scriptscriptstyle\rm T})
\end{array} \right) \left( 
\begin{array}{c}
A_{m} \\ B_{m}
\end{array}
\right) =\left( 
\begin{array}{c}
0 \\ 0
\end{array}
\right) .
\end{equation}
Non-trivial solutions exist if the determinant vanishes, and hereby
the wave vector $\kappa _{m}$ is a solution to the equation

\begin{equation}
\gamma ^{-1}=I_{m}(\kappa _{m}R_{\scriptscriptstyle\rm T})
K_{m}(\kappa _{m}R_{\scriptscriptstyle\rm T}),  
\label{kappa_m}
\end{equation}
where the result for the Wronskian $W\left\{ K_{m}\left( x\right)
  ;I_{m}\left( x\right) \right\} =1/x$ has been used \cite{OLIVER}.
Expanding Eq. (\ref {kappa_m}) in the small-$\kappa
_{m}R_{\scriptscriptstyle\rm T}$ limit \cite{OLIVER} we find that a
bound state with angular momentum $m\hbar $ exists for $\gamma >2m$.
The number of bound states for a certain value of $\gamma $ is given
by $N={\rm Int}(\gamma /2)+1$ where ${\rm Int}(x)$ is the integer part
of $x$.  Thus, there is always at least a single bound state
corresponding to $m=0$.

From the matching conditions and the normalization of $R_{m}$ (see
Appendix \ref{normalization}), it follows that
\begin{eqnarray}
R_{m}(r) &=&A_{m}\times  \left\{\begin{array}{ccc} 
I_{m}(\kappa _{m}r) &,& r<R_{\scriptscriptstyle\rm T}\\
\frac{I_{m}(\kappa_{m}R_{\scriptscriptstyle\rm T})}
{K_{m}(\kappa _{m}R_{\scriptscriptstyle\rm T})}K_{m}(\kappa _{m}r)
&,&R_{\scriptscriptstyle\rm T}<r\end{array}\right. ,
\label{wavefunction_exact} \\
A_{m} &=&\frac{\sqrt{2}}{R_{\scriptscriptstyle\rm T}}
\left[ \frac{I_{m}^{2}(\kappa _{m}R_{\scriptscriptstyle\rm T})}
{K_{m}^{2}(\kappa _{m}R_{\scriptscriptstyle\rm T})}
K_{m-1}(\kappa _{m}R_{\scriptscriptstyle\rm T})
K_{m+1}(\kappa_{m}R_{\scriptscriptstyle\rm T})
-I_{m-1}(\kappa _{m}R_{\scriptscriptstyle\rm T})
I_{m+1}(\kappa _{m}R_{\scriptscriptstyle\rm T})\right] ^{-1/2}.  
\label{A}
\end{eqnarray}
A plot of the radial wave function is shown in the inset of Fig. 3.
Increasing the confinement strength, the radial wave function becomes
more localized which is accompanied by an increase in the binding
energy.

\subsection{Metal contacts}

For the metal it is convenient to assume a cylindrical geometry and
consider a cylindrical wire of radius $R_{\scriptscriptstyle\rm C}\gg
R_{\scriptscriptstyle\rm T}$. The Hamiltonian is written as

\begin{equation}
\hat{{\cal {H}}}_{\scriptscriptstyle\rm C} =
-\frac{\hbar ^{2}}{2m_e}\left[ \frac{\partial ^{2}}{\partial z^{2}}
+\frac{\partial ^{2}}{\partial r^{2}}
+\frac{1}{r}\frac{\partial }{\partial r}
+\frac{1}{r^{2}}\frac{\partial ^{2}}{\partial \phi ^{2}}\right]
+V_{\scriptscriptstyle\rm C}(r), 
\end{equation}
with a hard-wall confining potential 
\begin{equation}
V_{\scriptscriptstyle\rm C}(r)=\left\{ 
\begin{array}{ccc}
-V_{\scriptscriptstyle\rm C}^0 & , & r<R_{\scriptscriptstyle\rm C} \\ 
\infty  & , & R_{\scriptscriptstyle\rm C} < r
\end{array}\right. .
\end{equation}

Obviously, for $r\geq R_{\scriptscriptstyle\rm C}$, $\Psi (r,\phi
,z)=0$ and for $r<R_{\scriptscriptstyle\rm C}$ the eigenstates have
the form
\begin{eqnarray}
\Psi _{\nu m}(r,\phi ,z) &=&R_{\nu m}(r)\chi _{m}(\phi )\psi _{\nu m}(z), \\
R_{\nu m}(r) &=&C_{\nu m}J_{m}(\kappa _{\nu m}r), \label{R_contact} \\
\chi _{m}(\phi ) &=&(2\pi )^{-1/2}\exp [im\phi ], \\
\psi _{\nu m}(z) &=&\left[ k_{\nu m}(E)\right] ^{-1/2}
\exp \left[ \pm ik_{\nu m}(E)z\right] ,
\end{eqnarray}
where $J_{m}$ is a Bessel function of the first kind of order $m$,
$\kappa _{\nu m}^{2}=2m_e\varepsilon _{\nu m}/\hbar ^{2}$ is a wave
vector corresponding to the radial energy $\varepsilon_{\nu m}$,
$k_{\nu m}(E)=\left[ \frac{2m_e}{\hbar ^{2}}\left(
    E+V_{\scriptscriptstyle\rm C}^0-\varepsilon_{\nu m}\right) \right]
^{1/2}$ is the wave vector of the longitudinal motion, and $E$ is the
total energy of the state. Again the normalization $\left[ k_{\nu
    m}(E)\right] ^{-1/2}$ makes the propagating modes carry the same
amount of current.

The boundary condition for the radial wave function leads to
$J_{m}(\kappa _{\nu m}R_{\scriptscriptstyle\rm C})=0,$ from which we
find $\kappa_{\nu m}$ numerically. Since $J_{m}(x)\sim (2/\pi
x)^{1/2}\cos (x-m\pi /2-\pi /4)$ for large $x$ \cite{OLIVER}, we have
$\kappa _{\nu m}R_{\scriptscriptstyle\rm C}\simeq (\nu +m/2-1/4)\pi $
with $\nu =1,2,3,\ldots $.  The normalization $C_{\nu m}$ is given by
(see Appendix \ref{normalization})

\begin{equation}
C_{\nu m}=\frac{1}{R_{\scriptscriptstyle\rm C}}
\sqrt{\frac{2}{
-J_{m-1}(\kappa _{\nu m}R_{\scriptscriptstyle\rm C})
J_{m+1}(\kappa _{\nu m}R_{\scriptscriptstyle\rm C})}},  
\label{C}
\end{equation}
which is a real number.

\section{Transmission of contact}

We consider an electron in the tube in mode $m$ incident on the
contact (see Fig. \ref{FIG1}) and compute the transmission and
reflection coefficients. We construct the scattering states in the
basis of the eigenstates of the Schr\"{o}dinger equation (see previous
section). Since the
angular momentum is a conserved quantity, the transmitted and
reflected parts of the wave function also have the same quantum number
$m$. In the quantum tube ($z<0$) the scattering state is given by

\begin{equation}
\Psi_{m}(r,\phi ,z) =R_{m}(r)\frac{\exp (im\phi )}{\sqrt{2\pi }}
\left[\frac{\exp (ik_{m}z)}{\sqrt{k_{m}}} +r_{m} 
\frac{\exp(-ik_{m}z)}{\sqrt{k_{m}}}\right], 
\end{equation}
and in the contact ($z>0$) by 
\begin{equation}
\Psi_{m}(r,\phi ,z)=\sum_{\nu =1}^{\infty }t_{\nu m}R_{\nu m}(r)
\frac{\exp(im\phi )}{\sqrt{2\pi }}
\frac{\exp (ik_{\nu m}z)}{\sqrt{k_{\nu m}}}. \nonumber
\end{equation}
Here $r_{m}$ is the reflection amplitude for mode $m$ and $t_{\nu m}$
is the corresponding transmission amplitude. We assume that the
effective electron mass is the same in the two materials so that the
continuity of $\Psi_m (r,\phi ,z)$ and $\partial \Psi_m (r,\phi
,z)/\partial z$ at $z=0$ are appropriate boundary conditions. For
carbon nanotubes and metals like Al and Au this is a reasonable
approximation ($m_e^\star\sim 1.2\, m_0$ \cite{DRESSELHAUS}). For
general details on how to account for differences in the effective
mass and the underlying symmetry of the lattice we refer to Refs.
\cite{BASTARD} and references therein. The boundary conditions lead to

\begin{eqnarray}
r_{m} &=&\frac{1-\sum_{\nu =1}^{\infty }\varrho _{\nu m}^{2}}
{1+\sum_{\nu =1}^{\infty }\varrho _{\nu m}^{2}}, \\
t_{\nu m} &=&\frac{2\varrho _{\nu m}}
{1+\sum_{\nu =1}^{\infty }\varrho _{\nu m}^{2}},
\end{eqnarray}
where $\varrho _{\nu m}\equiv \sqrt{k_{\nu m}/k_{m}}\left\langle
  R_{m}\left| R_{\nu m}\right. \right\rangle$, with the radial overlap
defined as $\left\langle R_{m}\left| R_{\nu m}\right. \right\rangle
\equiv \int_{0}^{\infty }{\rm d}r\,rR_{m}(r)R_{\nu m}(r).$ In addition
we have the sum-rule $\sum_{\nu =1}^{\infty }\left\langle R_{m}\left|
    R_{\nu m}\right. \right\rangle ^{2}=1$, which can be used to verify
the numerical convergence. The overlap can be calculated analytically
(see Appendix \ref{overlap}) and the squared overlap is given by

\begin{eqnarray}
\left<R_m\left|R_{\nu m}\right.\right>^2 &=& 
\left(\frac{R_{\scriptscriptstyle\rm T}}
{R_{\scriptscriptstyle\rm C}}\right)^2
\frac{\left[
J_m(\kappa_{\nu m}R_{\scriptscriptstyle\rm T}) 
+ \kappa_{\nu m} I_m(\kappa_mR_{\scriptscriptstyle\rm T} )
K_m(\kappa_mR_{\scriptscriptstyle\rm C}) 
J_{m+1}(\kappa_{\nu m }R_{\scriptscriptstyle\rm C})\right]^2}
{\left[\left(\kappa_mR_{\scriptscriptstyle\rm T}\right)^2 
+\left(\kappa_{\nu m}R_{\scriptscriptstyle\rm T}\right)^2\right]^2 
\left[J_{m-1}(\kappa_{\nu m}R_{\scriptscriptstyle\rm C})
J_{m+1}(\kappa_{\nu m}R_{\scriptscriptstyle\rm C})\right]}\\
&\quad&\times
\frac{4}{ \left[K_m^2(\kappa_m R_{\scriptscriptstyle\rm T})
I_{m-1}(\kappa_m R_{\scriptscriptstyle\rm T}) 
I_{m+1}(\kappa_m R_{\scriptscriptstyle\rm T}) 
-I_m^2(\kappa_m R_{\scriptscriptstyle\rm T})
K_{m-1}(\kappa_m R_{\scriptscriptstyle\rm T}) 
K_{m+1}(\kappa_m R_{\scriptscriptstyle\rm T})\right]}  \nonumber.
\end{eqnarray}
The total transmission from mode $m$ in the quantum tube into the contact is
thus

\begin{eqnarray}
{\cal T}_{m} &=&\sum_{\nu =1}^{\infty }{\cal {P}}_{\nu m}
\left| t_{\nu m}\right| ^{2}  \nonumber \\
&=&\left| \frac{2}
{1+\sum_{\nu =1}^{\infty }\varrho _{\nu m}^{2}}\right|^{2}
\sum_{\nu =1}^{\infty }{\cal {P}}_{\nu m}\varrho _{\nu m}^{2},
\end{eqnarray}
where ${\cal {P}}_{\nu m}$ projects onto the propagating modes
($k_{\nu m}$ real) of the metal contact. Here we have assumed that the
lengths of the quantum tube and the contact are semi-infinite so that
tunneling through evanescent modes can be neglected. These should be
included in the case of two metal contacts connected by a quantum tube
of finite length. Introducing real and imaginary parts by $\sum_{\nu
  =1}^{\infty }\varrho _{\nu m}^{2}\equiv {\Gamma }_{m}^{\prime
  }+i{\Gamma }_{m}^{\prime \prime }$ we obtain

\begin{equation}
{\cal T}_{m}=\frac{4{\Gamma }_{m}^{\prime }}
{\left( 1+{\Gamma }_{m}^{\prime }\right) ^{2}
+{{\Gamma }_{m}^{\prime \prime }}^{2}}.  
\label{T_m}
\end{equation}
The reflection probability ${\cal R}_m$ can be calculated in a similar
manner which provides us with the usual sum-rule ${\cal T}_{m}+{\cal
  R}_{m}=1$, ensuring the conservation of probability current density.

To summarize, the transmission probability of mode $m$ can be
calculated
from Eq. (\ref{T_m}) with $k_{m}(E)=\left[ \frac{2m_e}{\hbar ^{2}}%
  \left( E-\varepsilon _{m}\right) \right] ^{1/2}$ and $k_{\nu
  m}(E)=\left[
\frac{2m_e}{\hbar ^{2}}\left( E+V_{\scriptscriptstyle\rm C}^0-\varepsilon _{\nu m}\right) %
\right] ^{1/2}$, with $\varepsilon _{m}=\frac{\hbar ^{2}}{2m_e}\kappa
_{m}^{2}$ being the energy of the $m$th transverse mode in the tube
and similarly $\varepsilon _{\nu m}=\frac{\hbar ^{2}}{2m_e}\kappa
_{\nu m}^{2}$ is the transverse energy in the contact. Here $\kappa
_{m}$ and $\kappa _{\nu m}$ are solutions to $I_{m}(\kappa
_{m}R_{\scriptscriptstyle\rm T})K_{m}(\kappa
_{m}R_{\scriptscriptstyle\rm T})=\gamma ^{-1}$ and $J_{m}(\kappa _{\nu
  m}R_{\scriptscriptstyle\rm C})=0$, respectively. For a numerical
implementation, an upper cut-off $\nu _{c}$ in the sum over modes in
the contact is needed and the sum-rule for the squared radial overlap
is then a measure of the numerical convergence for a given cut-off.

When choosing values for the confinement parameters, $\gamma $ and
$V_{\scriptscriptstyle\rm C}^0$, we take into account that the Fermi
momenta of the quantum tube and the metal can be different since the
relevant electrons are located at the Fermi levels of the two
materials. For the metal contact we use the known Fermi energies for
e.g. Al and Au to relate the confinement potential to the Fermi level
as $E_{\rm F}^{\scriptscriptstyle\rm C}=E+V_{\scriptscriptstyle\rm
  C}^0$, with the Fermi energy being defined positive. For the tube,
the Fermi energy enters as $E_{{\rm F}}^{\scriptscriptstyle\rm
  T}=E+|\varepsilon _{m}|$. When the two materials are brought into
contact the chemical potentials align, but the difference in Fermi
wave vectors remains. Thus for the metal contact
$E+V_{\scriptscriptstyle\rm C}^0=\hbar^2 \left[k_{{\rm
      F}}^{\scriptscriptstyle\rm C}\right]^{2}/2m_e$, and for the tube
$E+|\varepsilon _{m}|=\hbar^2\left[ k_{{\rm F}}^{\scriptscriptstyle\rm
    T}\right]^{2}/2m_e$. For the tube we need to specify $\gamma$,
which follows from the work function $W=|\varepsilon _{m}|-\hbar^2
\left[k_{{\rm F}}^{\scriptscriptstyle\rm T}\right]^{2}/2m_e$. We have
neglected the charge density induced at the interface by a mismatch of
the work functions. For a discussion of this in the context of the
screening properties of one-dimensional systems, see e.g. Ref.
\cite{ODINTSOV}.

\section{Contact resistance of single-walled armchair carbon nanotubes}

\begin{figure}
\begin{center}
\epsfig{file=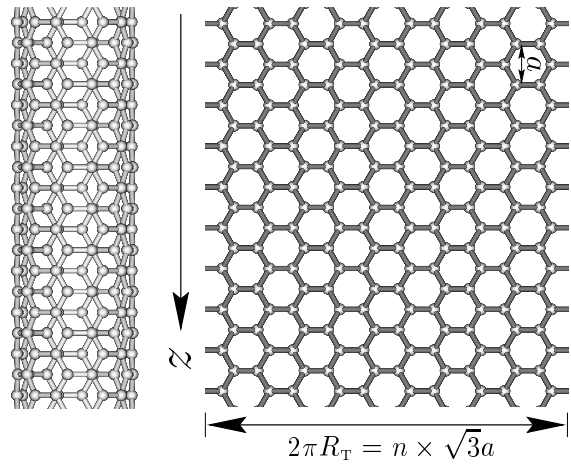, width=0.4\columnwidth,angle=-90}
\end{center}
\caption{A sheet of graphite (upper panel) which can be rolled up to a
  $(n,n)$ armchair single-walled carbon nanotube (lower panel). The
  tube has a periodicity $a\simeq 0.246\,{\rm nm}$ along the tube axis
  ($z$-axis) and a radius $R_{\scriptscriptstyle\rm T}\simeq n\times
  \sqrt{3}a/2\protect\pi $ with $4n$ carbon atoms along the perimeter,
  arranged in two rows that resemble a chain of armchairs. The shown
  example is a $(5,5)$ armchair nanotube.}
\label{FIG2}
\end{figure}

The $(n,n)$ armchair single-walled carbon nanotube can be regarded as
the result of rolling one sheet of graphite (with the carbon atoms in
a hexagonal lattice) in the direction of one of the bonds
\cite{DRESSELHAUS}.  The resulting tube has a periodicity $a\simeq
0.246\,{\rm nm}$ along the tube axis ($z$-axis) and a radius
$R_{\scriptscriptstyle\rm T}\simeq n\times \sqrt{3}a/2\pi$ with $4n$
atoms along the perimeter, arranged in two rows that resemble a chain
of armchairs, see Fig. \ref{FIG2}. Their metallic character is
caused by two $\pi$ bands crossing the Fermi level at a wave vector
$k_{{\rm F}}^{\scriptscriptstyle\rm T} \simeq 2\pi/3a$.

\begin{figure}
\begin{center}
\epsfig{file=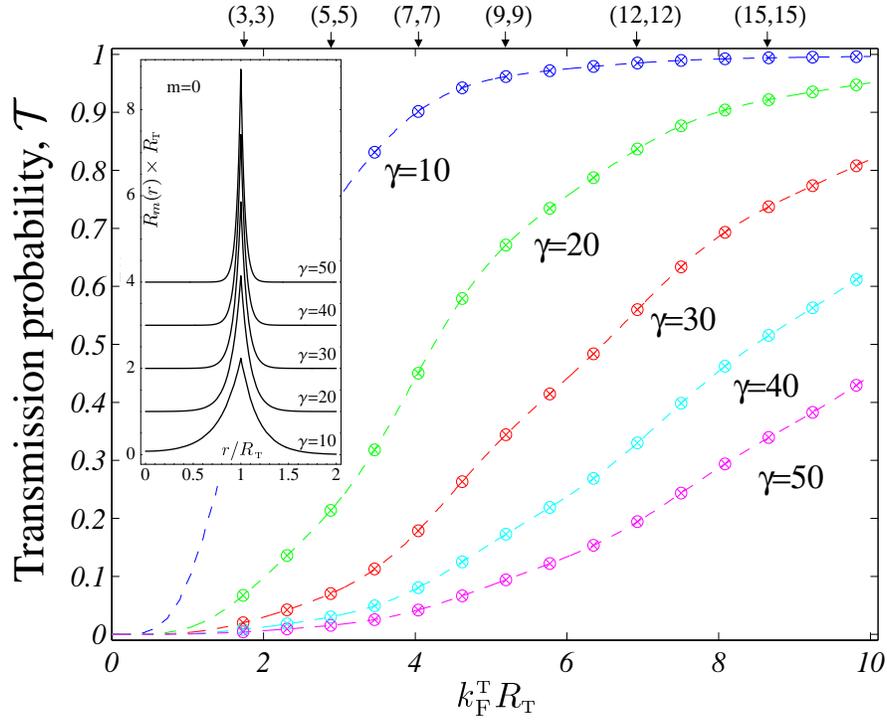, width=0.8\columnwidth}
\end{center}
\caption{Transmission probability ${\cal T}$ from an $(n,n)$ armchair
  carbon nanotube of radius $R_{\scriptscriptstyle\rm T}\simeq
  n/\sqrt{3}k_{{\rm F}}$ into a cylindrical contact with radius
  $R_{\scriptscriptstyle\rm C}\gg R_{\scriptscriptstyle\rm T}$. ${\cal
    T}$ is shown for several values of the confinement strength
  $\protect\gamma$. The results marked with $\otimes$ are the specific
  values of $R_{\scriptscriptstyle\rm T} k_{{\rm F}}=n/\sqrt{3}$
  corresponding to $(n,n)$ tubes and the dashed lines are calculated
  curves which are shown as guides to the eye. The inset shows the
  radial wave function of the quantum tube, Eq.
  (\ref{wavefunction_exact}), for $m=0$. The nanotube and the metal
  are assumed to have the same Fermi wave vectors, so that $k_{{\rm
      F}}^{\scriptscriptstyle {\rm T}}/k_{{\rm F}}^{\scriptscriptstyle
    {\rm C}}=1$.}
\label{FIG3}  
\end{figure}

\begin{figure}[htb]
\begin{center}
\epsfig{file=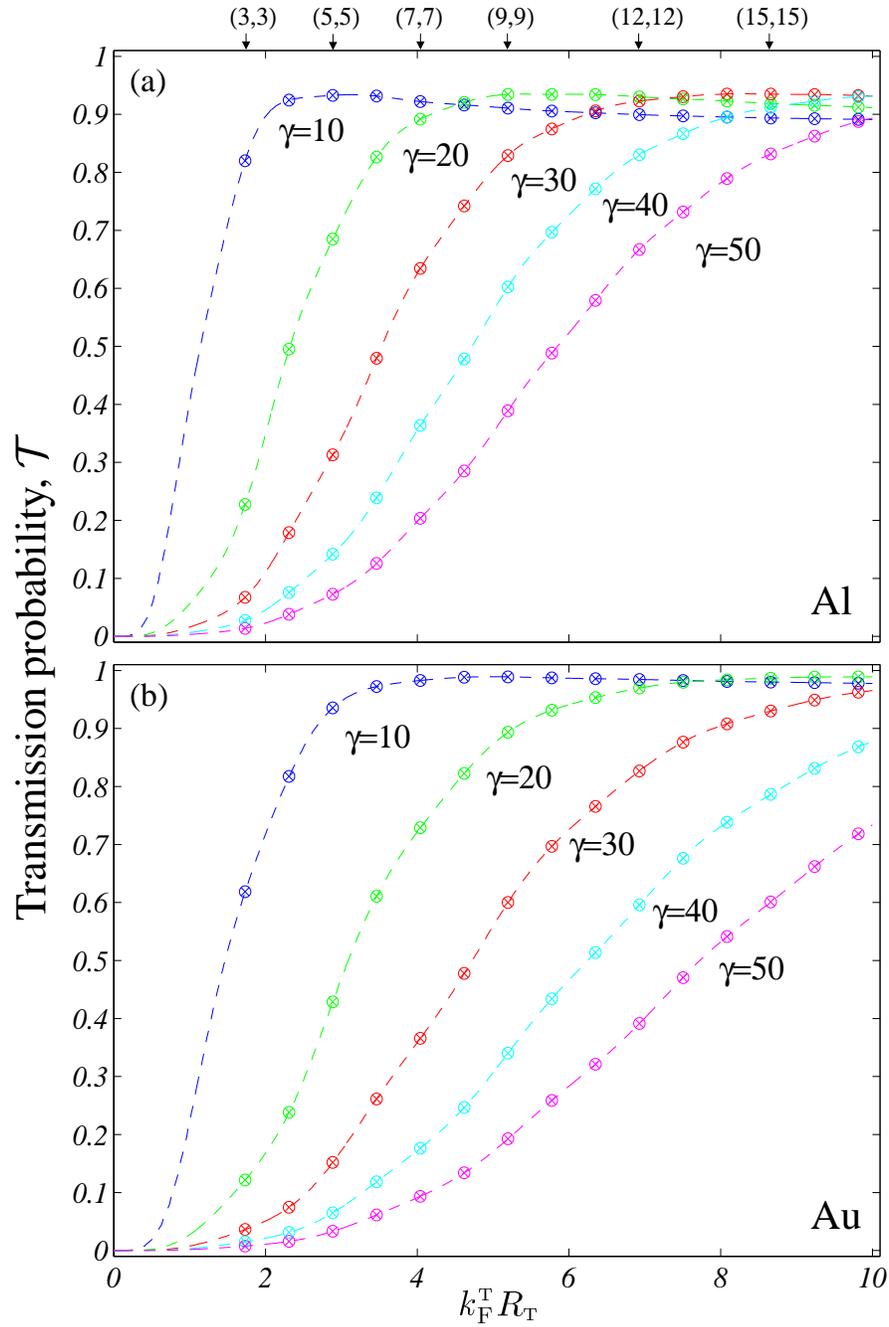, width=0.8\columnwidth}
\end{center}
\caption{The same calculation as in Fig. \ref{FIG3}, but, with a mismatch
  between the Fermi wave vectors of the carbon nanotube and the metal.
  Panel (a) is for $k_{{\rm F}}^{\scriptscriptstyle {\rm T}}/k_{{\rm
      F}}^{ \scriptscriptstyle {\rm Al}}\sim 0.49$ corresponding to
  nanotubes embedded in an Al contact and panel (b) is for $k_{{\rm
      F}}^{\scriptscriptstyle {\rm T}}/k_{{\rm F}}^{
    \scriptscriptstyle {\rm Au}}\sim 0.70$ corresponding to nanotubes
  embedded in an Au contact.}
\label{FIG4}
\end{figure}

As discussed recently by Tersoff \cite{TERSOFF99} the metallic
armchair carbon nanotubes have electrons at the Fermi level which can
be regarded as having an angular momentum quantum number $m=0$. In
order to apply our simple model to the problem of the contact
resistance of $(n,n)$ single-walled carbon
nanotubes embedded in a free-electron metal we notice that $k_{{\rm F}}^{%
  \scriptscriptstyle \rm T}R_{\scriptscriptstyle\rm T}\simeq
n/\sqrt{3}$. In Fig. \ref{FIG3} the transmission probability at the
Fermi level is shown for several values of $k_{\rm
  F}^{\scriptscriptstyle\rm T} R_{\scriptscriptstyle\rm T}$
corresponding to $(n,n)$ armchair carbon nanotubes for various values
of the dimensionless confinement strength $\gamma$.  In the particular
case of an Al contact, the mismatch is given by $k_{{\rm
    F}}^{\scriptscriptstyle\rm T}/k_{{\rm F}}^{\scriptscriptstyle {\rm
    Al}}\sim 0.49$ and the corresponding transmission is presented in
panel (a) of Fig. \ref{FIG4}. In panel (b) we show similar results for
an Au contact for which $k_{{\rm F}}^{\scriptscriptstyle\rm T}/k_{{\rm
    F}}^{\scriptscriptstyle {\rm Au}}\sim 0.70$.

\begin{figure}[htb]
\begin{center}
\epsfig{file=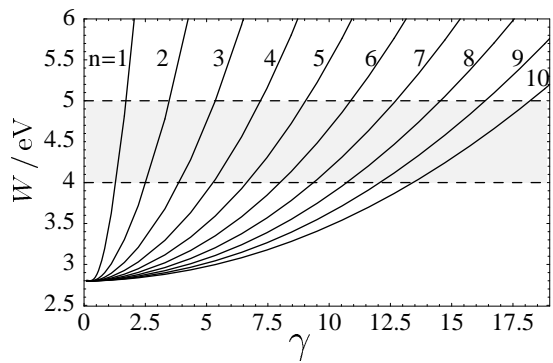, width=0.49\columnwidth}
\end{center}
\caption{Work function $W$ of quantum tubes corresponding to $(n,n)$
  armchair carbon nanotube as a function of the dimensionless
  confinement strength $\gamma$. Experimentally, the work function of
  a carbon nanotube is $W\sim 4-5\, {\rm eV}$ which for the shown
  nanotubes means that $\gamma <10-20$ is the important parameter
  range.}
\label{FIG5}
\end{figure}

In order to estimate $\gamma$ we relate it to the work function of the
carbon nanotube which is of the order $4-5\,{\rm eV}$.  For the
quantum tube we associate a work function to the $m=0$ bound state via
its binding energy, i.e.  $W\equiv
\frac{\hbar^2\kappa_0^2}{2m_e}-E_{{\rm F}}^{\scriptscriptstyle\rm T}$
where $\kappa_0$ is the solution to $I_0(\kappa_0
R_{\scriptscriptstyle\rm T})K_0(\kappa_0 R_{\scriptscriptstyle\rm
  T})=\gamma^{-1}$. In Fig. \ref{FIG5} this work function is shown as
a function of the confinement strength for quantum tubes corresponding
to $(n,n)$ armchair carbon nanotubes. From this we estimate that
$\gamma < 10-20$ is a reasonable regime for the curves shown in Figs.
\ref{FIG3} and \ref{FIG4}. This means that a transmission close to
unity (per band at the Fermi level) can be expected if the armchair
nanotube is embedded in free-electron metals like Al or Au. Assuming a
work function $W=4.5\,{\rm eV}$ of the nanotube we have used the
corresponding value of $\gamma$ to calculate the transmission for the
different nanotubes. For $(n,n)$ armchair nanotubes with Al and Au
contacts we find ${\cal T}_{\scriptscriptstyle\rm Al}\sim 0.93$ and
${\cal T}_{\scriptscriptstyle\rm Au}\sim 0.98$, respectively. In the
case of matching Fermi wave vectors ($k_{{\rm F}}^{\scriptscriptstyle
  {\rm T}}/k_{{\rm F}}^{\scriptscriptstyle {\rm C}}=1$) we get ${\cal
  T}\sim 0.87$. For the considered tubes ($3 \leq n\leq 17$), these
transmissions are found to be almost independent of the specific value
of the tube indices $(n,n)$.

Here we have only taken the geometry-related contact scattering into
account. Physically, a lower transmission can be caused by electrons
being scattered by interface imperfections/roughness, deviations from
a spherical Fermi surface of the metal contact, and scattering due to
non-matching work functions of the nanotube and metal. Also scattering
due to the non-matching Fermi velocities of the nanotube and the metal
could be expected. However, as shown in Fig.  \ref {FIG4}, a mismatch
between Fermi wave vectors can actually in some cases increase the
transmission (and thereby the conductance) due to quantum
interferences, even though the mismatch by itself is known to give
rise to momentum relaxation and thereby resistance.

\section{Discussion and conclusion}

We have considered the contact resistance (in terms of transmission)
of a quantum tube embedded in a free-electron metal. For the quantum
tube we have modeled the radial confinement of the electron motion by
an attractive delta function potential which gives rise to at least
one bound state in the radial direction. The strength of the
attractive potential can phenomenologically be associated to the work
function of the quantum tube. Within this model we have calculated the
transmission of a quantum tube contacted by a free-electron metal. Due
to the cylindrical geometry of the contact, considerable analytical
progress was possible and with the resulting equations the scattering
problem is readily solved numerically.

As an application we have considered the transparency of contacts with
armchair carbon nanotubes embedded in free-electron metals.  Our
calculations show that in the absence of scattering mechanisms
associated to e.g. interface imperfections/roughness, deviations from
a spherical Fermi surface of the metal contact, and scattering due to
non-matching work functions of the nanotube and metal, the geometry
itself allows for a high transparent contact between armchair carbon
nanotubes and free-electron metal contacts.  Furthermore, from this
simple model we find that Al would be a good candidate for such a
metal as it was suggested recently by Tersoff \cite{TERSOFF99}.  For
Au however, we find that the present 3D geometry allows for good
contact in contrast to Tersoff's findings for Au, which were based on
1D considerations.

\section*{Acknowledgements}

We would like to thank M. Brandbyge, H. Bruus, D.H. Cobden, and J.
Nyg\aa rd for useful discussions.

\appendix

\section{Normalization of radial wave functions}

\label{normalization}

From the radial wave function of the quantum tube, Eq. (\ref
{wavefunction_exact}), it follows that the normalization is given by

\begin{equation}
A_m= \kappa_m \left[\int_0^{\kappa_m R_{\scriptscriptstyle\rm T}}
{\rm d}\alpha\,\alpha I_m^2(\alpha)+ 
\frac{I_m^2(\kappa_mR_{\scriptscriptstyle\rm T})}
{K_m^2(\kappa_mR_{\scriptscriptstyle\rm T})}
\int_{\kappa_mR_{\scriptscriptstyle\rm T}}^{\infty}
{\rm d} \alpha\,\alpha K_m^2(\alpha)\right]^{-1/2},
\end{equation}
and since \cite{LUKE}

\begin{eqnarray}
\int {\rm d}\alpha\,\alpha I_m^2(\alpha)&=& 
{\alpha}^2\left[I_m^2({\alpha})-I_{m-1}({\alpha})I_{m+1}
({\alpha})\right]/2, \\
\int{\rm d}\alpha\,\alpha K_m^2(\alpha)&=& {\alpha}^2
\left[K_m^2({\alpha})-K_{m-1}({\alpha})K_{m+1}({\alpha})\right]/2,
\end{eqnarray}
we get the result in Eq. (\ref{A}). Similarly, from the radial wave function
of the free-electron metal contact, Eq. (\ref{R_contact}), it follows that
the normalization is given by

\begin{equation}
C_{\nu m}= \left[\int_0^{R_{\scriptscriptstyle\rm C}}
{\rm d}r\,r J_m^2(\kappa_{\nu m}r)\right]^{-1/2},
\end{equation}
and since $J_m(\kappa_{\nu m}R_{\scriptscriptstyle\rm C})=0$ and \cite{LUKE}

\begin{equation}
\int{\rm d} \alpha\, \alpha J_m^2(\alpha)= \alpha^2
\left[J_m^2(\alpha)-J_{m-1}(\alpha)J_{m+1}(\alpha)\right]/2,
\end{equation}
we obtain Eq. (\ref{C}).

\section{Overlap of radial wave functions}

\label{overlap}

The overlap of radial wave functions can be written as

\begin{eqnarray}
\left<R_m\left|R_{\nu m}\right.\right>&=& 
A_m C_{\nu m}
\left[\int_0^{R_{\scriptscriptstyle\rm T}}{\rm d} r \, 
r I_m(\kappa_m r)J_m(\kappa_{\nu m}r) 
+ \frac{I_m(\kappa_m R_{\scriptscriptstyle\rm T})}
{K_m(\kappa_m R_{\scriptscriptstyle\rm T})} 
\int_{R_{\scriptscriptstyle\rm T}}^{R_{\scriptscriptstyle\rm C}}{\rm d}r \, 
r K_m(\kappa_m r) J_m(\kappa_{\nu m}r) \right]  \nonumber \\
&=& \frac{A_m C_{\nu m}}{\kappa_m^2
+\kappa_{\nu m}^2} \frac{ J_m(\kappa_{\nu m}R_{\scriptscriptstyle\rm T}) 
+\kappa_{\nu m}R_{\scriptscriptstyle\rm C}
I_m(\kappa_mR_{\scriptscriptstyle\rm T})
K_m(\kappa_mR_{\scriptscriptstyle\rm C})
J_{m+1}(\kappa_{\nu m }R_{\scriptscriptstyle\rm C})} 
{K_m(\kappa_m R_{\scriptscriptstyle\rm T})},
\end{eqnarray}
where we have used the integrals \cite{LUKE}

\begin{eqnarray}
\int{\rm d}r\, r I_m(\alpha r) J_m(\beta r)&=& 
\frac{r\left\{\alpha I_{m+1}(\alpha r)J_m(\beta r) 
+ \beta I_m(\alpha r) J_{m+1}(\beta r)\right\}}{\alpha^2+\beta^2}, \\
\int{\rm d}r\, r K_m(\alpha r) J_m(\beta r)&=& 
-\frac{r\left\{\alpha K_{m+1}(\alpha r)J_m(\beta r) 
- \beta K_m(\alpha r) J_{m+1}(\beta r)\right\}}{\alpha^2+\beta^2},
\end{eqnarray}
together with the boundary condition $R_{\nu
  m}(R_{\scriptscriptstyle\rm C})=0$.

\end{document}